\documentclass{aa}  

\usepackage{graphicx}
\usepackage{txfonts}
\def\Teff{$T\rm_{eff}$}
\def\kms{$\mathrm{km\, s^{-1}}$}

\def\Vt{$V_{\rm t}$}
\def\logg{$\log\,g$}
\def\vsini{{\sl v~sin~i~}}

\def\loggf{$\log\,gf$}
\def\Rg{$R_{G}$}

\def\kms{km s$^{-1}$}

\def\FeH{$\mathrm{[Fe/H]}$}

\def\II{\,{\sc ii}}
\def\I{\,{\sc i}}
\def\III{\,{\sc iii}}

\begin{document}

\title{Unique distant classical Cepheid OGLE GD-CEP-1353 with anomalously high abundances of s- and r-process elements}

   \author 
          {
          V. V. Kovtyukh\inst{1,2}
          \and
          S. M. Andrievsky\inst{1,2,3}
          \and
          K. Werner\inst{2}
          \and
          S. A. Korotin\inst{4}}

 \institute{Astronomical Observatory, Odessa National University, 
             Shevchenko Park, 65014, Odessa, Ukraine\\
              \email{vkovtyukh@ukr.net}
         \and
Institut f\"{u}r Astronomie und Astrophysik, Kepler Center for 
Astro and Particle Physics, Universit\"{a}t T\"{u}bingen, 
Sand 1, 72076 T\"{u}bingen, Germany
         \and
GEPI, Observatoire de Paris, Universit\'e PSL, CNRS, 
5 Place Jules Janssen, F-92190 Meudon, France
         \and
Physics of stars department, Crimean Astrophysical Observatory, 
Nauchny 298409, Republic of Crimea
             }

\date{Received date; accepted: date}

\abstract
{}
{While looking for recently discovered distant Cepheids with an interesting
chemical composition, we noticed one star (OGLE GD-CEP-1353) with 
extremely large equivalent widths of spectral lines of heavy elements. 
The aim of this work is to perform an abundance analysis,
and to find a possible explanation for the found chemical anomaly.}
{Quantitative analysis of the equivalent widths and synthetic spectrum
synthesis were used to derive abundances in this star. Both local and nonlocal 
thermodynamic equilibrium (LTE and NLTE) approximations were used in our analysis.}
{Abundances of 28 chemical elements from carbon to thorium were derived.
While light and iron peak elements show abundances typical for distant Cepheids
(located in the outer disk), the s-process elements
are overabundant about one dex. r-process elements are slightly
less overabundant. This makes the star a unique Cepheid of our
Galaxy.}
{}

\keywords{Stars: abundances -- Stars: variables: Cepheids}

\authorrunning{Kovtyukh et al.}
\titlerunning{Unique Cepheid in the Galaxy}

\maketitle

\section{Introduction}

Cepheids are yellow supergiant stars of spectral classes F to K, which are
at an evolutionary stage when they are crossing the instability strip in the 
Hertzsprung-Russell diagram (HRD). During the crossing time, these stars pulsate 
in radial modes.

Observationally, their atmospheres do not exhibit remarkable chemical peculiarities.
A few anomalies that are an inherent feature of Cepheids are connected to their
internal structure. After hydrogen depletion in the stellar core, it contracts
and heats. A shell source, where fresh hydrogen is converted into helium, 
develops. At this evolutionary stage, the star expands and settles in the red giant 
branch (RGB) region. Here the first dredge-up occurs (\citealt{Iben1967}). To the 
surface of the star, it brings material processed in the incomplete CNO cycle. Here, carbon is 
partially converted into nitrogen, and if the first dredge-up happens before nitrogen
is converted into oxygen, one can detect that the surface abundances of carbon 
and nitrogen are altered: carbon is depleted and nitrogen is enhanced
(see, for example, the recent large-scale study of the Cepheids' chemical properties
by \citealt{Luck2018}, Fig. 19). This phenomenon is not observed in the F and G 
unevolved dwarfs (e.g., \citealt{Reddyetal2003}).
The next stage after the RGB evolution is determined by the
helium burning in the stellar core. As the helium in the core ignites in a
smooth regime, the star contracts and increases its effective temperature.
Thus, the star starts to perform a blue loop in the HRD. At this
time interval, the star can be noticed as a Cepheid. This scenario is well
known, and its description can be found in many literature sources.
What is important is that the Cepheid atmospheres preserve the C--N abundance
anomalies. This is a well-established observational fact (see, for
example, \citealt{Lambert1981}, \citealt{LuckLambert1985}, \citealt{Luck2018} 
and references therein). Cepheid stars show a slightly increased abundance of 
sodium (e.g., \citealt{Andrievskyetal2003}, \citealt{Luck2018}). Elements that 
are formed through the capture of free thermalized neutrons by seed nuclei
(s-process elements) normally do not show significantly increased abundances, and this 
is clear from the evolutionary point of view. The internal structure of 
Cepheids does not provide a neutron source. Nearly normal relative
abundances of the neutron capture elements ([s/Fe]) in Cepheids were reported 
recently by \cite{daSilvaetal2016} and \cite{Luck2018}. As a rule, all of the
studied heavy element abundances, including r-process
elements such as Eu, are in the range from zero to 0.2 dex ({s/Fe}). 
For zirconium and lanthanum, abundances are 0.1--0.2 dex higher but, according to 
\cite{Luck2018}, with a big error bar. 

In summary, any anomalies of heavy element abundances that can be found in a 
Cepheid deserve special attention. In this paper we report our detection of 
remarkable overabundances of s- and r-process elements in the distant classical Cepheid 
OGLE GD-CEP-1353. Its characteristics are given in Table~\ref{parameters} (together with atmospheric
parameters described in the next Section). The high-resolution spectrum of this star was obtained 
using the Ultraviolet and Visual Echelle Spectrograph (UVES; \citealt{Dekker2000}) at 
the Very Large Telescope of ESO at Paranal (Chile). The resolving power and signal-to-noise ratio 
(S/N) are 42300 and 34, respectively.
%%%\LEt{ Please remember to spell out acronyms upon first appearance in the abstract and then again beginning with the introduction. 
%%%Thereafter the acronym should be used unless it’s at the beginning of a sentence.***} 

%Table 1
\begin{table*}
\caption{Some characteristics of our program Cepheid OGLE GD-CEP-1353.}
\label{parameters}
\centering
\begin{tabular}{ccccccccccccc}
\hline\hline            
     P      & JD  & phase& <V>   &  <I>  & \Teff& $\sigma$(\Teff) &\logg&  \Vt&  $l$    &  $b$    & d & \Rg   \\ 
     d      & 2\,400\,000+ &      & mag   &  mag  &   K  &  K&    & \kms&       &       &  kpc      & kpc  \\ 
\hline  

  3.1492138 & 59198.63235  & .839 & 15.501& 13.676&  5817&$\pm$ 137& 2.00& 3.30&247.481&--4.282&  11.0 & 16.0 \\     
\hline  
\hline
\end{tabular}
\end{table*}

\cite{Skowronetal2019} give the heliocentric (d) and Galactocentric (R$_{\rm G}$) distances that 
we list in Table~1. 
%%%\LEt{ A\&&A discourages authors from directly addressing the reader - for example "note that" can be either deleted completely or replaced with "we note that"”.***}
We note that from the Gaia DR3 (Data Release 3) database OGLE GD-CEP-1353 has a parallax of 0.0815 mas, 
which means its heliocentric distance is about 12 kpc, being close to above the mentioned value.
%%%\LEt{ Please introduce.***} 

% Fig1 
\begin{figure}
 \resizebox{\hsize}{!}{\includegraphics{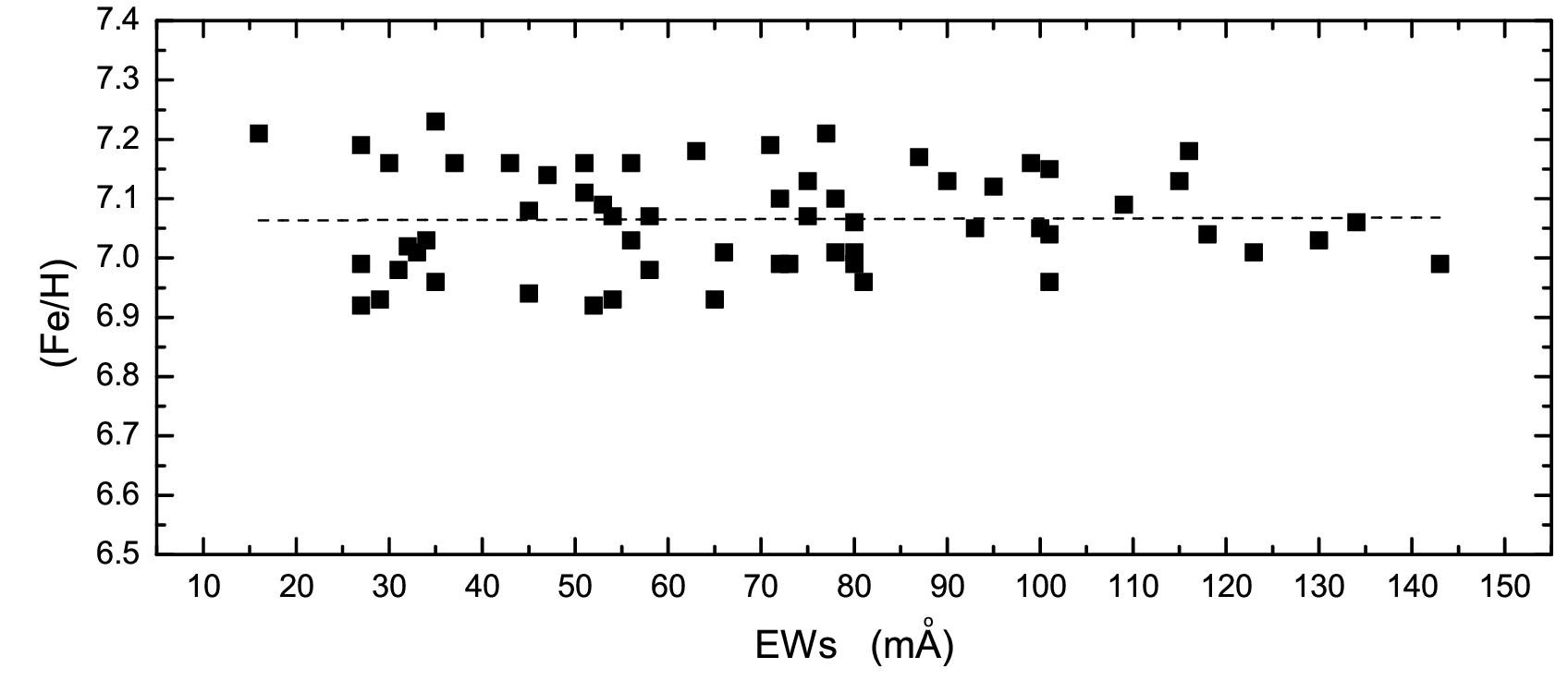}}
 \caption{To the \Vt\ determination (absence of a dependence between the iron abundance of individual 
 lines and their equivalent widths).}
 \label{Vt}
 \end{figure}  

The remainder of this paper is organized as follows. In Sect.\,2 we  
describe our abundance analysis of OGLE GD-CEP-1353.
In Sect.\,3 we discuss our results and present our conclusions.

\section{Abundance results}

The first attempt to derive abundances in this star was recently made by \cite{Trentinetal2023} 
(ASASSN-V J074354.86-323013.7). Their paper lists abundance results for 24 elements from carbon to 
neodymium for a large number of Cepheids (abundances of only 18 elements were determined for this  
star; from heavy species only for zirconium and barium). Unfortunately, this unique star was apparently 
"lost" in the large number of the program stars, and the authors did not pay due attention and did not 
discuss its individual chemical properties.

\subsection{Atmosphere parameters}

The effective temperature, \Teff, was derived from the line-depth ratios \citep{Kovtyukh2007}, 
a technique commonly employed in studies of Cepheid variables \citep[e.g.,][]{Andrievskyetal2016,
Luck2018,Lemasleetal2013,daSilvaetal2022,Kovtyukhetal2022}. Once \Teff\ is determined, the surface gravity 
(\logg) is found by imposing the iron ionization balance (the same iron abundance derived from the lines of 
neutral and ionized iron). The microturbulent velocity, \Vt, was derived assuming that there is 
no dependence between the iron abundance, obtained from Fe~{\sc i} lines, 
%%%\LEt{ Depending on the meaning, please write "or the equivalent..." or "and that equivalent widths (EWs) are of...".***}
and the equivalent widths (EWs) of the same lines (Fig.~\ref{Vt}). The adopted value [Fe/H] is that, which is derived from the Fe~{\sc i} 
lines, since we assumed the ionization balance and because they outnumber Fe~{\sc ii} lines. 
The atmospheric parameters \Teff, \logg,\ and \Vt\ are listed in Table~\ref{parameters}.
Since \cite{Trentinetal2023} used practically the same methods of the atmosphere parameters' determination, 
we have very close results on the temperature, gravity, and microturbulent velocity.

\subsection{LTE results}

The abundances of different elements were derived in the LTE approximation using atmosphere models interpolated 
for the   atmosphere parameters within the grid of {\sc ATLAS9} models  by \cite{CastelliKurucz2004}. 
We discarded strong lines (with EWs$>$150 m\AA) due to noticeable damping effects. The list of the lines of the
heavy elements measured in the spectrum of our program star is given in Table \ref{EWs} of the Appendix.
%; the full version of this table is available online. 
The oscillator strengths, \loggf, were adopted from the Vienna Atomic Line Database 
({\sc VALD}, \citealt{Ryabchikovaetal2015}, version 2023). The reference solar abundances were taken from \cite{Asplundetal2009} 
or determined by us in the NLTE calculations. 

\begin{figure*}
 \resizebox{\hsize}{!}{\includegraphics{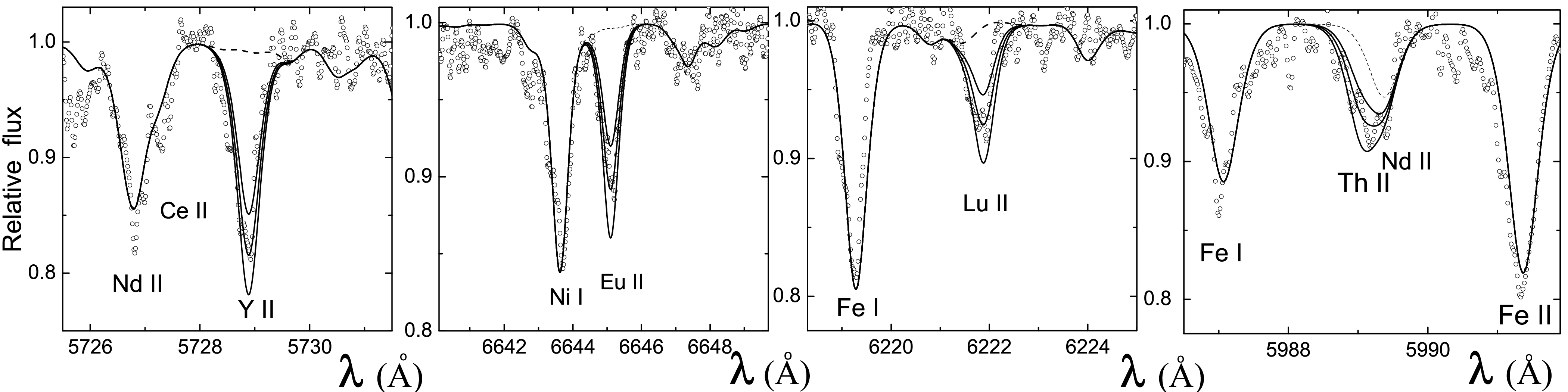}}
 \caption{Observed (open circles) and LTE synthetic profiles of Y\II\ 5728.89 \AA, Eu\II\ 6645.10 \AA, 
 Lu\II\ 6221.89 \AA, and Th\II\ 5989.045 \AA\ lines (solid line). The dashed line indicates no Y, Eu, Lu, or Th. 
 The solid lines show the abundance variation of $\pm$ 0.20 dex for these elements and the best fit.}
 \label{synth}
 \end{figure*}

\begin{figure*}
 \resizebox{\hsize}{!}{\includegraphics{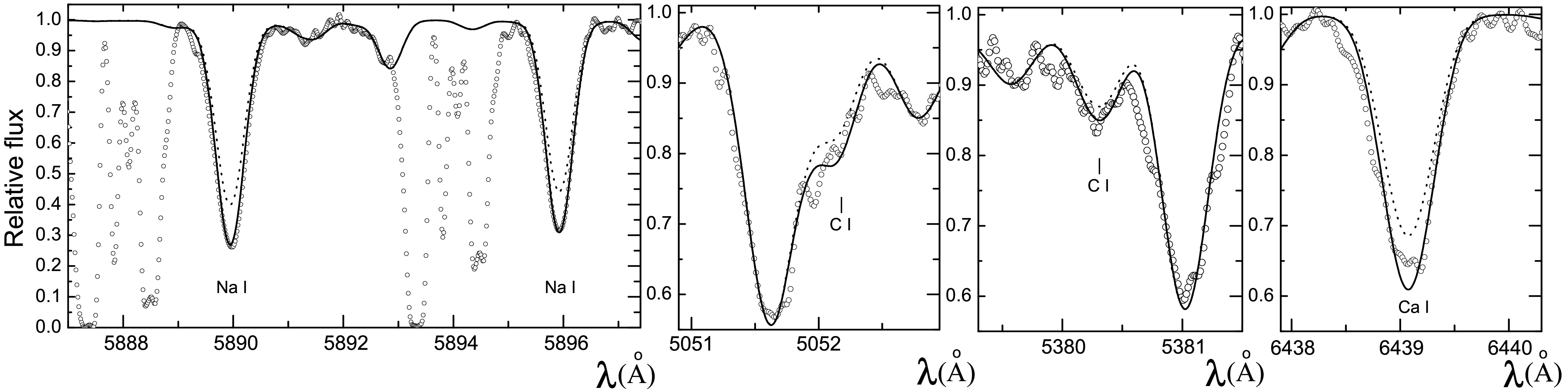}}
 \caption{NLTE profiles of some lines of C, Na, and Ca (solid line) compared to the observed 
 profiles (open circles). LTE profiles were calculated with the same abundances as NLTE
 profiles (dotted line).}
 \label{profile}
 \end{figure*}

The abundances of some heavy elements were calculated via direct fitting observed and synthetic profiles of individual 
spectral lines. We used the SynthV code of \cite{Tsymbaletal2019} in combination with the ATLAS9 model.
LTE approximation was assumed in this spectrum synthesis. Some examples of synthetic spectra fragments are shown 
in Fig.~\ref{synth} and \ref{profile}. All of the results are given in Table~\ref{abundance}. 
 
% Fig2 

\subsection{NLTE results}

For some elements we applied the NLTE approximation to derive their abundances.
In Table~2 the corresponding ions are marked as NLTE. The atomic
models used are described in detail in several papers by the members of our group, 
%%%\LEt{ e.g./i.e. should be written out in full when part of the main text (not inside parentheses or figure legends). e.g. can be replaced with ‘for example,’ ‘for instance, 
%%% or ’such as’ when in the main text. i.e. should be replaced by ‘that is’ or similar when in main text. See the entries for e.g. and i.e. in Section 2.1, ‘Note 2’ of the language %%%guide.***}
for example, carbon (\citealt{Andrievskyetal2001}, \citealt{Lyubimkovetal2015}), sodium
(\citealt{KorotinMishenina1999}, \citealt{Dobrovolskasetal2014}), magnesium
(\citealt{Misheninaetal2004}, \citealt{{Cerniauskasetal2017}}), aluminum
(\citealt{Andrievskyetal2008}, \citealt{Caffauetal2019}), sulfur (\citealt{Korotin2009},
calcium (\citealt{Andrievskyetal2018}), and barium (\citealt{Andrievskyetal2009}). 

The general method of the NLTE calculations we used in our previous papers is
the following. In order to find atomic level populations for the ions
of interest, we employed the code {\sc MULTI} (\citealt{Carlsson1986}).
%%% \LEt{ Please introduce if possible.***}. 
For our aim, this program was modified by \cite{Korotin1999}. 
MULTI allows one to calculate a single NLTE line profile. If the line 
of interest is blended, we performed the following procedure. With the help
of~{\sc MULTI}, we first calculated
%%%\LEt{ A\&&A uses the past tense to describe specific methods used in a paper, and the present tense to describe general methods and the findings of recent papers. For more details, %%%see Sect. 6 of the language guide https://www.aanda.org/for-authors/language-editing/6-verb-tenses. Please review my edits to ensure this was carried out appropriately throughout %%%your paper.***} 
the departure coefficients for those atomic levels 
that are responsible for the formation of the considered line. Then we included 
these coefficients in the LTE synthetic spectrum code {\sc SynthV} (\citealt{Tsymbaletal2019}). 
This allowed us to calculate the source function and opacity for each studied line. 
Simultaneously, the blending lines were calculated in LTE with the help of the line list
and corresponding atomic data from the {\sc VALD} database (\citealt{Ryabchikovaetal2015}) 
in the wavelength range of the line under study. For all our computations, we 
used 1D LTE atmosphere models computed with the {\sc ATLAS9} code by \cite{CastelliKurucz2004}.

%Fig. 3                                                               

Table~\ref{abundance} shows the averaged individual chemical abundances and their uncertainties. 
We note that abundance results may suffer from a combined influence of the errors in \Teff, \logg, \Vt, 
the position of the continuum, and the accuracy of the \loggf. Abundances of those elements that are 
represented in the spectrum by less than three lines must be taken with caution.

\begin{table*}
\caption{LTE and NLTE abundances in OGLE GD-CEP-1353.}
\label{abundance}
\centering
\begin{tabular}{rllcrrcrl}
\hline\hline            
 Code&  Ion& (El/H)&   $\sigma$& NL &  [El/H]&      Sun& [El/Fe] &        Remark\\  
\hline                       
 6.00&  C\I   & 8.29  &    0.06  &   2&  --0.27&   8.56  &   0.24  &              \\
 6.00&  C\I   & 8.00  &    0.12  &   3&  --0.43&   8.43  &   0.08  &       NLTE   \\
 8.00&  O\I   & 8.45  &    0.15  &   1&  --0.26&   8.71  &   0.25  &       NLTE   \\
11.00&  Na\I  & 6.39  &    --    &   1&   +0.07&   6.32  &   0.58  &              \\
11.00&  Na\I  & 6.04  &     0.12 &   4&  --0.21&   6.25  &   0.30  &       NLTE   \\
12.00&  Mg\I  & 7.48  &     --   &   1&  --0.20&   7.68  &   0.31  &              \\
12.00&  Mg\I  & 7.55  &    0.12  &   3&   +0.01&   7.54  &   0.52  &       NLTE   \\
13.00&  Al\I  & 6.34  &    0.20  &   4&  --0.09&   6.43  &   0.42  &       NLTE   \\
14.00&  Si\I  & 7.19  &    0.27  &   7&  --0.41&   7.60  &   0.10  &              \\
14.01&  Si\II & 7.39  &    0.16  &   2&  --0.21&   7.60  &   0.30  &              \\
16.00&  S\I   & 7.09  &    --    &   1&  --0.12&   7.21  &   0.39  &              \\
16.00&  S\I   & 6.64  &    0.25  &   5&  --0.52&   7.16  & --0.01  &       NLTE   \\
20.00&  Ca\I  & 6.04  &    0.16  &   5&  --0.34&   6.38  &   0.17  &              \\
20.00&  Ca\I  & 5.92  &    0.10  &  13&  --0.39&   6.31  &   0.12  &       NLTE   \\
21.01&  Sc\II & 2.89  &    0.15  &   4&  --0.36&   3.25  &   0.15  &              \\
22.00&  Ti\I  & 4.69  &    0.27  &   3&  --0.34&   5.03  &   0.17  &              \\
22.01&  Ti\II & 4.53  &    0.27  &   3&  --0.50&   5.03  &   0.01  &              \\
23.00&  V\I   & 3.82  &    --    &   1&  --0.29&   4.11  &   0.22  &              \\
24.00&  Cr\I  & 5.16  &    0.14  &   2&  --0.60&   5.76  & --0.09  &              \\
24.01&  Cr\II & 5.26  &    0.09  &   3&  --0.50&   5.76  &   0.01  &              \\
25.00&  Mn\I  & 4.95  &    0.09  &   3&  --0.59&   5.54  & --0.08  &              \\
26.00&  Fe\I  & 7.07  &    0.09  &  60&  --0.51&   7.58  &   0.00  &              \\
26.01&  Fe\II & 7.08  &    0.10  &  10&  --0.50&   7.58  &   0.01  &              \\
28.00&  Ni\I  & 5.88  &    0.12  &   8&  --0.42&   6.30  &   0.09  &              \\
29.00&  Cu\I  & 4.19  &    0.15  &   4&  --0.06&   4.25  &   0.45  &       NLTE   \\
39.01&  Y\II  & 2.81  &    0.19  &  11&   +0.62&   2.19  &   1.13  &              \\
40.01&  Zr\II & 3.24  &    0.26  &   3&   +0.36&   2.88  &   0.87  &              \\
56.01&  Ba\II & 3.58  &    0.15  &   3&   +1.41&   2.17  &   1.92  &       NLTE   \\
57.01&  La\II & 2.18  &    0.16  &  16&   +0.85&   1.33  &   1.36  &              \\
58.01&  Ce\II & 2.46  &    0.09  &  18&   +0.75&   1.71  &   1.26  &              \\
59.01&  Pr\II & 1.26  &    0.17  &   6&   +0.45&   0.81  &   0.96  &              \\
60.01&  Nd\II & 2.18  &    0.13  &  35&   +0.66&   1.52  &   1.17  &              \\
60.02&  Nd\III & 2.12 &    0.20  &   1&   +0.60&   1.52  &   1.11  &              \\
62.01&  Sm\II & 1.82  &    0.20  &   6&   +0.72&   1.10  &   1.23  &              \\
63.01&  Eu\II & 0.76  &    0.24  &   1&   +0.09&   0.52  &   0.60  &      synth   \\
68.01&  Er\II & 1.95  &    0.20  &   1&   +1.03&   0.92  &   1.54  &      synth   \\
71.01&  Lu\II & 0.60  &    0.20  &   1&   +0.40&   0.10  &   0.91  &      synth   \\
90.01&  Th\II & 1.00  &    0.20  &   1&   +0.98&   0.02  &   1.49  &      synth   \\
\hline     
\end{tabular}
\\
Note 1: NL is the number of lines used.\\
Note 2: (El/H) is the absolute abundance value on the scale where the hydrogen abundance is 12.00.

\end{table*}

The program star abundance pattern that referred to solar abundances is also presented in Fig.~\ref{elements}. 
The extremely high barium abundance might not be realistic due to the very strong lines of this element.
These lines are effectively formed in the upper layers of the Cepheid atmosphere, where the 
microturbulent velocity can significantly exaggerate the value found from the iron lines. 
We could not account for this possible phenomenon, so we simply state that the barium abundance from 
our analysis may be overestimated. 

As mentioned, all derived abundances were normalized to the iron content in our program star.
It should be noted that the iron content in Cepheids shows the expected dependence upon Galactocentric 
distance: the larger the distance, the lower the iron content. For instance, this is shown 
in \cite{Andrievskyetal2016}, \cite{Luck2018}, \cite{Trentinetal2023}, \cite{daSilvaetal2023}, 
and the references therein. 
From these data, it can be seen that the distant Cepheids in the outer Galactic disk
have an iron content 2--3 times lower than in the Sun (see, for example, Fig.~23 in
\citealt{Luck2018} based on period--luminosity distances). The distance to our program
Cepheid is mentioned in Table 1. Cepheids at such a distance show absolute iron content 
from 7.0 to 7.4 (7.5 for the Sun), and thus our program star fits in this range.

\section{Discussion and conclusion}

First of all, it should be noted that none of the Cepheids of our Galaxy studied to date 
show such an abnormal chemical composition. In some sense, it resembles the chemical composition
of the chemically peculiar (CP) stars (\citealt{Ryabchikovaetal1997}, \citealt{Yushchenkoetal2008}), 
or even Przybylski's star (\citealt{Shulyaketal2010}), of course, in a much lesser extent. 
It is believed that CP stars gain their anomalies from atomic
diffusion in the dynamically stable atmospheres (\citealt{Michaud1973}). This cannot 
be the case for yellow supergiants with atmospheric convection. As to Przybylski's star,
several hypotheses exist explaining its extreme peculiarities (for more details, readers 
can refer to the overview in \citealt{Andrievsky2022}).
%%%\LEt{ Including long stretches of text within parentheses should be avoided in English, and thus my proposed edits in the next paragraph. In this particular case, you might %%%consider writing "For more details, readers can refer to the overview in...".***} 
In addition, this author considered processes in a binary system consisting 
of Przybylski's star and a neutron star, which is the source of high-energy
$\gamma$ radiation, which may affect the atmosphere of Przybylski's star.

\begin{figure}
 \resizebox{\hsize}{!}{\includegraphics{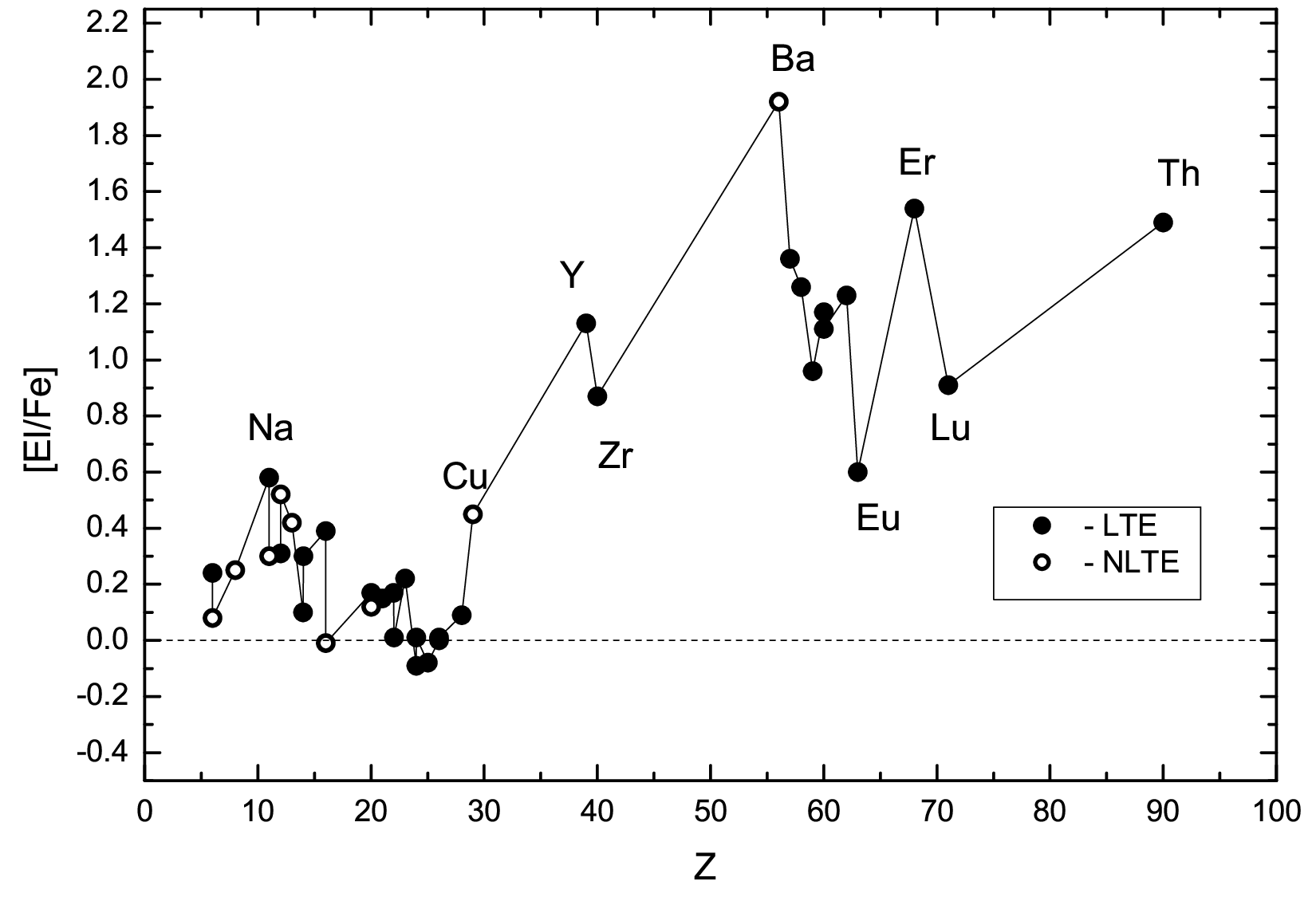}}
 \caption{Graphical representation of the content of Table~\ref{abundance}.}
 \label{elements}
 \end{figure}

Studying the derived abundance distribution in our program star, we must describe a few 
characteristic features of it (Fig.~\ref{elements}), namely, the deficiency of iron-peak elements; the rather 
high relative-to-iron abundance of carbon; the increased sodium abundance, for example for the Cepheids with 
a pulsational period of 3 days, \citealt{Genovalietal2015} give an average sodium overabundance [Na/Fe]
of about 0.3 dex, while our program star shows [Na/Fe] = 0.6 dex; and a significantly 
increased abundance of the s-process elements (elements that formed as a result of the slow neutron 
capture by seed nuclei). In addition, the following remarks can be made: there is a fairly high level of
overabundance of the light s-process elements (Y, Zr); and with europium, lutetium, and thorium, being 
r-process elements (rapid neutron capture), they have apparently high abundances, too. We note that
according to \cite{daSilvaetal2016} (their Fig. 6), with the relative-to-iron europium abundance [Eu/Fe] being 
extrapolated to a distance of about 12 kpc, it is about zero, while the corresponding value for our
program star is 0.6 dex.

As we mentioned in the Introduction, Cepheid atmospheres exhibit the results of the dredge-up event.
From the observational point of view, the result of the first dredge-up event was described by \cite{Lambert1981} and 
\cite{LuckLambert1985}. After dredge-up carbon becomes deficient, while nitrogen is overabundant. The region 
of the spectrum of our program star that is available to us does not contain nitrogen lines, so we cannot prove this fact. 
Yellow supergiant stars and Cepheids, in particular, show a moderate overabundance of sodium 
(\citealt{Andrievskyetal2003}), which can be a sign of the neon-sodium cycle operation. Our NLTE calculations 
of the sodium abundance in this star confirm this fact. However, increased sodium in our program Cepheid may 
have another origin.

% Fig. 4                                                              

Nearly normal relative abundances of the neutron capture elements ([s/Fe]) in Cepheids were recently reported 
by \cite{daSilvaetal2016} and \cite{Luck2018}. As it was already mentioned in the Introduction, all studied 
s-process elements normally do not show significant excesses in the relative-to-iron abundances. If we compare our results 
on s-process abundances presented in Table~2 with those given in the two abovementioned papers, we would see that
Cepheid OGLE GD-CEP-1353 has an overabundance of these elements by several times larger than any
studied Cepheid until now (more than 500 stars). It makes this object really unique in our Galaxy. 
Below we consider several possible hypotheses that may help to understand its unique chemical properties.

Our program star OGLE GD-CEP-1353 has a light curve typical of classical Cepheids (\citealt{Udalski2015}).
Additionally, its Galactic latitude is small (Table~\ref{parameters}), which means it can be a thin disk star.
Since its fundamental pulsational period is about 3.15 days, according to \cite{Turner1996} 
(mass--pulsation period relation), its mass should be about 4 solar masses. The lifetime of a star of this 
mass (B-star progenitor + following a Cepheid stage), 
$t \approx 10^{10} (\frac{M_{\odot}}{M_{*}})^{2.5}$ years, should be about $3.7 \times 10^{8}$ years (the core helium-burning stage 
takes about 25--30 \% of the main-sequence time,  see, e.g., \citealt{Chiosi1990}). According to the age--pulsation period 
relation by \cite{Bonoetal2005}, the age of a Cepheid is 1.1 $\times 10^{8}$ years (for \FeH =--0.5). 
 
1. Suppose now that this Cepheid had a companion of a slightly higher mass. After a certain time, such a companion
ends its evolution as an asymptotic giant branch (AGB) star. Having finished this evolutionary stage, the envelope 
of this rather massive AGB star (more than 4 solar masses) could contaminate the present Cepheid's atmosphere (or its B-star progenitor), 
enriching it in s-process elements. The more massive component after all becomes a C--O white dwarf. 
This scenario can explain the appearance of the s-process elements in the Cepheid's atmosphere. 
More massive AGB stars produce a larger amount of the weak s-process elements, such as 
%%\LEt{ A\&&A discourages the use of "like" since it can lead to ambiguity; "such as" and "similar to" are preferred alternatives depending on the context.***}
yttrium and zirconium, and this is seen in our program Cepheid (see, Fig.~\ref{elements}). 

According to \cite{GoswamiGoswami2023}, see their Fig.~14 Panel a,
%%%\LEt{ A\&A discourages the use of "see" in the main part of sentences, please consider writing "(2023; see their Fig.\ 14 Panel a)".***} 
the AGB model of 4 solar masses with a metallicity of about --0.5 (recall that the iron peak elements in our program star show the same abundance level) 
yield enough the s-process elements, and the light s-process elements (Sr, Y, and Zr) are at the 
same abundance level as the heavy s-process elements (Ba--Sm), [El/Fe] $\approx 0.8-0.9$. This is exactly 
what we see in our program Cepheid. It should be noted that in the lower-mass AGB stars, the light s-process 
elements are less abundant compared to the heavy s-process elements.  Finally, as is mentioned above, 
sodium is apparently increased in our program star. This may be connected to the fact that the highest sodium 
production is reached in 4 solar mass AGB stars (\citealt{DiCriscienzo2016}).  

However, despite its attractivity, this hypothesis faces a problem with unexpectedly high abundances of 
the typical r-process elements, such as europium, lutetium, and thorium. As believed, the nuclei of these 
heavy elements cannot be formed inside the AGB star 
%%%\LEt{ Please check the meaning hasn't changed; this sentence was not grammatically correct before.***}
because of the insufficient neutron flux. The typical neutron density in the low-mass AGB stars is about 
$10^{7}$~cm$^{-3}$ (\citealt{Stranieroetal2006}).

2. This Cepheid has a companion of a significantly higher mass. In this case its companion ended
its evolution as a neutron star after the 
%%%\LEt{ Please introduce.***}
Supernova~II (SN~II) explosion. For this to happen, the B-star progenitor
must have a mass larger than 8 solar masses. If it was a close explosion, the possibility of 
a single contamination event would be possible. However, there is one weak place in this hypothesis. 
Even for the lower limit of the SN~II mass, the expected lifetime of its progenitor B star is
about $6 \times 10^{7}$ years. Soon after this time, the star explodes. Its companion (presently a
Cepheid) is still evolving as a late B star. If the atmosphere of this B star is contaminated, 
then after the following evolution to the red giant stage and first dredge-up, all of the material
synthesized in the SN explosion that contaminated the B star atmosphere will be mixed with the
stellar interior. Thus, the surface abundance anomalies will hardly survive.

3. Finally, suppose that the OGLE GD-CEP-1353's more massive companion ended its evolution as 
a neutron star. If this neutron star is a source of $\gamma$ radiation and if this radiation is directed 
toward the present-day Cepheid atmosphere, then one can expect that irradiation of the atmosphere gas
will produce free neutrons. Such free neutrons (provided their flux is sufficiently high)
can lead to the production of the s- and r-process elements. A similar mechanism was
proposed by \cite{Andrievsky2022} to explain the phenomenon of a chemically peculiar
Przybylski's star (see details in that paper). 

An additional question may arise concerning the first and third scenario.
%%%\LEt{ A\&&A discourages authors from directly addressing the reader - for example question marks should be removed and sentences should be rephrased.***}
How long can the contaminated atmosphere of the Cepheid star preserve the chemical composition abundance anomalies? 
To estimate the characteristic time, we used the following formula (\citealt{Sweet1950}):
%%%\LEt{ Equations in the main text should be followed by a comma or a full stop when
%%%necessary, as if they were phrases forming part of a sentence.***}
$t = 8\times10^{12} \frac{M^{3}}{LR^{4}}\frac{1}{\Omega^{2}} ~~{\rm years.}$

All values here are in solar units. As previously mentioned, the mass of our program star with its pulsation 
period of about 3 days, is about 4 solar masses. According to \cite{Gieren1999}, a Cepheid with such 
a period has a radius of about 30 solar radii. The period -- luminosity relation gives the luminosity of this star 
of about $10^{3}$ $L_{\odot}$. To determine the angular rotation velocity of the Cepheid (the radii of the Cepheid and 
the Sun are known), we adopted a typical radius of its progenitor, the main-sequence B star, to be about 3.3 $R_{\odot}$, 
and a linear equatorial velocity about 200 \kms\ (a very rough estimate according to \citealt{Brottetal2011} gives 
\vsini\ of about 100 \kms\ for stars of about 10 solar masses, see their Fig.~1). Then, after expanding from 
3 to 30 $R_{\odot}$, the rotational velocity of the Cepheid decreased to a few \kms\ (approximately the same as 
the solar rotational rate). All adopted values give us a characteristic time of the large-scale meridional 
circulation of about $5 \times 10^{8}$ years or more. This value exceeds our program Cepheid lifetime 
(see the estimate above).

To evaluate the validity of the hypothesis described above, a necessary condition has to be met.
One needs to prove that OGLE GD-CEP-1353  is a component of a binary system with a secondary component being a compact 
object. Regarding the OGLE GD-CEP-1353 visual magnitude, it seems to be impossible now. 

In summary, we have described a very unusual chemical composition of this star, in particular the high abundance
of heavy elements in its atmosphere. It may be a component of a binary system originally consisting of two stars 
with slightly or moderately different masses. The further evolution of the  stars in this system led to contamination 
of the OGLE GD-CEP-1353 atmosphere with heavy elements, and thus forming in this way its unique chemical properties.  

\begin{acknowledgements}
 
Based on observations made with ESO Telescopes at the
Paranal Observatories under programme ID 0106.D-0561(A).
VVK and SMA are grateful to DAAD (German Academic Exchange Service) for financial support. 
SMA would also like to thank ESO at Garching for support during his stay in Germany,
which enabled him to perform part of this work.

We are grateful to our referee for his/her very detailed review and helpful comments.

\end{acknowledgements}

\clearpage

%%%%%%%%%%%%%%%%% APPENDICES %%%%%%%%%%%%%%%%%%%%%
%\newpage

\begin{appendix} 
\section{}

\onecolumn
\centering

\begin{longtable}{crrrrr}
\caption{Equivalent widths (in m\AA) of the lines used for the calculation of the abundances of elements (for LTE calculations we used lines with EW$<$150 m\AA\ only).} 
\label{EWs}\\
\hline 
%\hline
Wavelength   & Ion &  EW  &  \loggf&   Elow  &    (El/H)   \\
     (\AA)   &         &(m\AA)&        &    (eV) &             \\ 
\hline   
%\hline 
\endfirsthead \caption{continued.}\\ \hline\hline
Wavelength   & Ion &  EW  &  \loggf&   Elow  &    (El/H)   \\
     (\AA)   &       &(m\AA)&       &    (eV) &             \\   
\hline  
\endhead 
   5380.3370  & 6.00 & 69.8 &--1.615 &  7.685  &    8.33  \\
   6587.6100  & 6.00 & 47.4 &--1.021 &  8.537  &    8.25  \\
   6160.7470  &11.00 & 72.7 &--1.245 &  2.104  &    6.39  \\
   5711.0880  &12.00 &124.6 &--1.723 &  4.346  &    7.48  \\
   5948.5410  &14.00 & 79.8 &--1.130 &  5.082  &    7.10  \\
   6125.0210  &14.00 & 38.8 &--1.464 &  5.614  &    7.43  \\
   6155.1340  &14.00 & 62.8 &--0.754 &  5.619  &    7.03  \\
   6237.3190  &14.00 & 48.5 &--0.974 &  5.614  &    7.05  \\
   6244.4650  &14.00 & 31.8 &--1.090 &  5.616  &    6.94  \\
   6414.9800  &14.00 & 29.1 &--1.035 &  5.871  &    7.07  \\
   6721.8480  &14.00 & 38.1 &--1.526 &  5.863  &    7.70  \\   
   6347.1090  &14.01 &109.1 &  0.170 &  8.121  &    7.28  \\
   6371.3710  &14.01 &110.3 &--0.039 &  8.121  &    7.50  \\
   6748.7900  &16.00 & 36.0 &--0.529 &  7.868  &    7.09  \\
   5349.4650  &20.00 &123.6 &--0.309 &  2.709  &    6.10  \\
   5581.9650  &20.00 &122.1 &--0.554 &  2.523  &    6.13  \\
   5590.1140  &20.00 & 88.5 &--0.571 &  2.521  &    5.75  \\
   5601.2770  &20.00 &125.6 &--0.522 &  2.526  &    6.14  \\
   6166.4390  &20.00 & 66.6 &--1.142 &  2.521  &    6.06  \\
   5318.3490  &21.01 & 53.7 &--2.014 &  1.357  &    3.01  \\
   5667.1490  &21.01 &111.8 &--1.309 &  1.500  &    3.02  \\
   6245.6370  &21.01 &113.6 &--1.021 &  1.507  &    2.72  \\
   6604.6010  &21.01 &110.8 &--1.308 &  1.357  &    2.80  \\
   5210.3850  &22.00 &125.5 &--0.849 &  0.048  &    4.76  \\
   6258.7060  &22.00 & 74.1 &--0.239 &  1.460  &    4.92  \\
   6261.0970  &22.00 & 21.5 &--0.478 &  1.430  &    4.40  \\
   5211.5300  &22.01 &103.2 &--1.159 &  2.590  &    4.29  \\
   5268.6150  &22.01 &105.0 &--1.669 &  2.598  &    4.82  \\ 
   6606.9500  &22.01 & 31.5 &--2.789 &  2.061  &    4.49  \\
   6090.2140  &23.00 & 31.5 &--0.061 &  1.081  &    3.82  \\
   5329.1380  &24.00 & 52.6 &--0.007 &  2.914  &    5.06  \\
   5348.3150  &24.00 &127.6 &--1.210 &  1.004  &    5.26  \\
   5334.8690  &24.01 & 73.6 &--1.825 &  4.072  &    5.18  \\
   5420.9220  &24.01 & 59.6 &--2.457 &  3.758  &    5.35  \\
   5502.0670  &24.01 & 47.4 &--2.116 &  4.168  &    5.24  \\
   6013.5100  &25.00 & 59.0 &--0.351 &  3.072  &    5.00  \\
   6016.6700  &25.00 & 73.8 &--0.182 &  3.073  &    5.00  \\
   6021.8200  &25.00 & 70.8 &--0.053 &  3.075  &    4.84  \\
   4950.1060  &26.00 & 57.7 &--1.669 &  3.417  &    6.98  \\
   5029.6180  &26.00 & 34.1 &--2.049 &  3.415  &    7.03  \\
   5049.8200  &26.00 &190.4 &--1.354 &  2.279  &    7.21  \\
   5090.7740  &26.00 &108.8 &--0.399 &  4.256  &    7.09  \\
   5198.7110  &26.00 &123.3 &--2.134 &  2.223  &    7.01  \\
   5217.3890  &26.00 &134.1 &--1.074 &  3.211  &    7.06  \\
   5242.4910  &26.00 &115.1 &--0.967 &  3.634  &    7.13  \\
   5243.7770  &26.00 & 51.2 &--1.149 &  4.256  &    7.16  \\
   5339.9290  &26.00 &170.5 &--0.646 &  3.266  &    7.16  \\
   5373.7090  &26.00 & 52.5 &--0.859 &  4.473  &    7.09  \\
   5383.3690  &26.00 &192.4 &  0.645 &  4.313  &    7.17  \\
   5389.4790  &26.00 &100.9 &--0.409 &  4.415  &    7.15  \\
   5391.4590  &26.00 & 66.4 &--0.920 &  4.154  &    7.01  \\
   5398.2790  &26.00 & 71.9 &--0.669 &  4.446  &    7.10  \\
   5466.3960  &26.00 & 71.8 &--0.629 &  4.371  &    6.99  \\
   5554.8950  &26.00 & 79.8 &--0.439 &  4.549  &    7.06  \\
   5560.2120  &26.00 & 27.0 &--1.189 &  4.435  &    6.99  \\
   5565.7040  &26.00 & 93.4 &--0.212 &  4.608  &    7.05  \\
   5638.2620  &26.00 & 89.8 &--0.720 &  4.220  &    7.13  \\
   5679.0230  &26.00 & 46.7 &--0.820 &  4.652  &    7.14  \\
   5686.5300  &26.00 & 74.9 &--0.570 &  4.549  &    7.13  \\
   5731.7620  &26.00 & 34.5 &--1.200 &  4.256  &    6.96  \\
   5752.0320  &26.00 & 27.2 &--1.010 &  4.549  &    6.92  \\
   5753.1230  &26.00 & 78.1 &--0.687 &  4.260  &    7.01  \\
   5859.5860  &26.00 & 77.0 &--0.630 &  4.549  &    7.21  \\
   5862.3560  &26.00 & 98.9 &--0.330 &  4.549  &    7.16  \\
   5883.8170  &26.00 & 55.6 &--1.260 &  3.960  &    7.03  \\
   5984.8150  &26.00 & 80.3 &--0.195 &  4.733  &    6.99  \\
   5987.0650  &26.00 & 54.2 &--0.520 &  4.796  &    7.07  \\
   6003.0120  &26.00 & 72.7 &--1.100 &  3.882  &    6.99  \\
   6008.5560  &26.00 & 94.8 &--0.985 &  3.884  &    7.12  \\
   6027.0510  &26.00 & 52.1 &--1.089 &  4.076  &    6.92  \\
   6056.0050  &26.00 & 64.7 &--0.320 &  4.733  &    6.93  \\
   6065.4820  &26.00 &158.8 &--1.529 &  2.609  &    7.16  \\
   6151.6180  &26.00 & 32.8 &--3.295 &  2.176  &    7.01  \\
   6165.3600  &26.00 & 37.4 &--1.473 &  4.143  &    7.16  \\
   6173.3360  &26.00 & 54.4 &--2.880 &  2.223  &    6.93  \\
   6213.4300  &26.00 &101.0 &--2.481 &  2.223  &    7.04  \\
   6215.1440  &26.00 & 50.9 &--1.190 &  4.186  &    7.11  \\
   6219.2810  &26.00 &101.3 &--2.432 &  2.198  &    6.96  \\
   6229.2280  &26.00 & 32.3 &--2.670 &  2.845  &    7.02  \\
   6232.6410  &26.00 & 86.9 &--1.360 &  3.654  &    7.17  \\
   6240.6460  &26.00 & 29.4 &--3.230 &  2.223  &    6.93  \\
   6246.3190  &26.00 &130.4 &--0.771 &  3.603  &    7.03  \\
   6252.5550  &26.00 &167.3 &--1.699 &  2.404  &    7.21  \\
   6265.1340  &26.00 &100.4 &--2.550 &  2.176  &    7.05  \\
   6270.2250  &26.00 & 55.9 &--2.470 &  2.858  &    7.16  \\
   6322.6860  &26.00 & 74.9 &--2.430 &  2.588  &    7.07  \\
   6336.8240  &26.00 &118.4 &--0.852 &  3.686  &    7.04  \\
   6355.0290  &26.00 & 70.9 &--2.349 &  2.845  &    7.19  \\
   6358.6980  &26.00 & 57.7 &--4.340 &  0.859  &    7.07  \\
   6380.7430  &26.00 & 30.5 &--1.375 &  4.186  &    6.98  \\
   6408.0180  &26.00 &116.0 &--1.017 &  3.686  &    7.18  \\
   6419.9500  &26.00 & 80.1 &--0.239 &  4.733  &    7.01  \\
   6430.8460  &26.00 &159.6 &--2.005 &  2.176  &    7.17  \\
   6475.6240  &26.00 & 42.6 &--2.941 &  2.559  &    7.16  \\
   6481.8700  &26.00 & 63.4 &--2.981 &  2.279  &    7.18  \\
   6494.9810  &26.00 &207.8 &--1.268 &  2.404  &    7.26  \\
   6498.9390  &26.00 & 35.0 &--4.698 &  0.958  &    7.23  \\
   6518.3670  &26.00 & 45.3 &--2.560 &  2.832  &    7.08  \\
   6592.9140  &26.00 &143.4 &--1.472 &  2.728  &    6.99  \\
   6593.8710  &26.00 & 81.3 &--2.421 &  2.433  &    6.96  \\
   6597.5610  &26.00 & 27.1 &--1.069 &  4.796  &    7.19  \\
   6609.1100  &26.00 & 45.0 &--2.691 &  2.559  &    6.94  \\
   6627.5450  &26.00 & 16.2 &--1.590 &  4.549  &    7.21  \\
   6733.1510  &26.00 & 29.5 &--1.150 &  4.638  &    7.16  \\
   6750.1530  &26.00 & 77.8 &--2.618 &  2.424  &    7.10  \\
   4993.3580  &26.01 &123.3 &--3.639 &  2.807  &    7.13  \\
   5100.6640  &26.01 & 59.7 &--4.169 &  2.807  &    6.94  \\
   5256.9370  &26.01 & 67.2 &--4.181 &  2.891  &    7.11  \\
   5414.0730  &26.01 & 98.7 &--3.539 &  3.221  &    7.12  \\
   5627.4970  &26.01 & 39.6 &--4.129 &  3.387  &    7.18  \\
   5991.3760  &26.01 &115.6 &--3.539 &  3.153  &    7.22  \\
   6084.1110  &26.01 & 64.2 &--3.779 &  3.199  &    6.94  \\
   6113.3220  &26.01 & 50.7 &--4.109 &  3.221  &    7.13  \\
   6369.4620  &26.01 & 68.0 &--4.159 &  2.891  &    7.05  \\
   6416.9190  &26.01 &109.2 &--2.649 &  3.892  &    6.97  \\
   6446.4100  &26.01 & 40.7 &--1.959 &  6.223  &    7.68  \\
   5694.9830  &28.00 & 32.1 &--0.609 &  4.089  &    5.83  \\
   5754.6560  &28.00 & 50.3 &--2.329 &  1.935  &    5.73  \\
   6086.2810  &28.00 & 35.3 &--0.529 &  4.266  &    5.96  \\
   6108.1160  &28.00 & 58.2 &--2.600 &  1.676  &    5.83  \\
   6111.0700  &28.00 & 28.3 &--0.869 &  4.088  &    6.01  \\
   6186.7110  &28.00 & 25.4 &--0.959 &  4.105  &    6.06  \\
   6767.7720  &28.00 & 84.8 &--2.140 &  1.826  &    5.77  \\
   6772.3150  &28.00 & 35.3 &--0.979 &  3.658  &    5.81  \\
   4883.6821 & 39.01 &303.2&   0.070&   1.084 &     3.36    \\
   5087.4190 & 39.01 &248.1& --0.170&   1.084 &     3.11    \\
   5119.1120 & 39.01 &148.3& --1.359&   0.992 &     2.92    \\
   5200.4097 & 39.01 &243.2& --0.570&   0.992 &     3.33    \\
   5289.8166 & 39.01 & 64.6& --1.850&   1.032 &     2.50    \\
   5320.7831 & 39.01 & 34.9& --1.950&   1.084 &     2.27    \\
   5402.7742 & 39.01 &162.0& --0.630&   1.839 &     3.18    \\
   5473.3853 & 39.01 &131.3& --1.020&   1.738 &     3.07    \\
   5480.7303 & 39.01 &148.7& --0.990&   1.721 &     3.23    \\
   5509.8948 & 39.01 &204.8& --1.010&   0.992 &     3.22    \\
   5521.5500 & 39.01 & 98.6& --0.635&   1.738 &     2.63    \\
   5544.6114 & 39.01 &110.9& --1.090&   1.738 &     2.91    \\
   5546.0088 & 39.01 &127.0& --1.100&   1.748 &     3.11    \\
   5662.9241 & 39.01 &217.8&   0.200&   1.944 &     3.12    \\
   5728.8865 & 39.01 & 81.6& --1.120&   1.839 &     2.72    \\
   6613.7317 & 39.01 &111.2& --1.110&   1.748 &     2.87    \\
   6795.4156 & 39.01 &109.9& --0.913&   1.738 &     2.75    \\
   5350.3500 & 40.01 & 91.3& --1.160&   1.773 &     3.32    \\
   6100.1210 & 40.01 & 70.9& --1.500&   1.756 &     3.37    \\
   6114.8520 & 40.01 & 51.0& --1.400&   1.665 &     2.94    \\
   5853.6680 & 56.01 &369.2& --0.999&   0.604 &     3.88    \\
   6141.7130 & 56.01 &543.3& --0.075&   0.704 &     3.62    \\
   6496.8970 & 56.01 &502.0& --0.376&   0.604 &     3.68    \\
   4804.0400 & 57.01 &110.8& --1.490&   0.235 &     2.16    \\
   4809.0000 & 57.01 &119.1& --1.400&   0.235 &     2.17    \\
   5156.7300 & 57.01 & 86.9& --1.850&   0.126 &     2.12    \\
   5303.5300 & 57.01 &124.5& --1.350&   0.321 &     2.20    \\
   5377.0520 & 57.01 & 42.4& --0.430&   2.304 &     2.34    \\
   5381.9100 & 57.01 & 30.6& --0.720&   2.134 &     2.28    \\
   5482.2700 & 57.01 & 54.5& --2.230&   0.000 &     2.00    \\
   5671.5280 & 57.01 & 21.4& --0.870&   2.210 &     2.31    \\
   5863.6900 & 57.01 & 77.5& --1.370&   0.927 &     2.29    \\
   5936.2100 & 57.01 & 56.6& --2.060&   0.173 &     2.00    \\
   6126.0750 & 57.01 & 52.0& --1.240&   1.252 &     2.19    \\
   6129.5560 & 57.01 & 87.9& --1.500&   0.772 &     2.36    \\
   6172.7210 & 57.01 & 53.7& --2.320&   0.126 &     2.16    \\
   6320.3760 & 57.01 &131.3& --1.610&   0.173 &     2.28    \\
   6390.4800 & 57.01 &138.2& --1.410&   0.321 &     2.30    \\
   6671.4040 & 57.01 & 51.4& --2.030&   0.403 &     2.09    \\
   5037.7996 & 58.01 & 82.0& --0.570&   1.008 &     2.48    \\
   5117.1690 & 58.01 & 76.9& --0.049&   1.402 &     2.30    \\
   5117.9455 & 58.01 & 37.2& --0.810&   1.247 &     2.43    \\
   5386.7730 & 58.01 & 48.5& --0.710&   1.212 &     2.43    \\
   5435.2435 & 58.01 & 42.6& --2.000&   0.000 &     2.42    \\
   5468.3710 & 58.01 & 78.2& --0.070&   1.402 &     2.30    \\
   5472.2791 & 58.01 &112.8& --0.100&   1.247 &     2.54    \\
   5516.0816 & 58.01 & 31.6& --0.570&   1.615 &     2.43    \\
   5518.4889 & 58.01 & 57.8& --0.650&   1.155 &     2.41    \\
   5561.4450 & 58.01 & 26.5& --0.880&   1.458 &     2.49    \\
   5613.6940 & 58.01 & 44.7& --0.640&   1.420 &     2.50    \\
   5623.0011 & 58.01 & 39.9& --1.150&   0.956 &     2.48    \\
   5695.8474 & 58.01 & 35.1& --0.610&   1.626 &     2.53    \\
   5933.5816 & 58.01 & 43.4& --1.770&   0.327 &     2.50    \\
   5941.5447 & 58.01 & 49.5& --0.940&   1.107 &     2.53    \\
   5959.6878 & 58.01 & 31.7& --0.690&   1.626 &     2.54    \\
   6043.3730 & 58.01 & 84.9& --0.480&   1.206 &     2.55    \\
   6051.8154 & 58.01 & 63.5& --1.530&   0.232 &     2.39    \\
   6272.0254 & 58.01 & 54.7& --0.400&   1.544 &     2.47    \\
   6652.7390 & 58.01 & 33.5& --0.580&   1.528 &     2.33    \\
   5034.4060 & 59.01 & 28.7& --0.090&   1.111 &     1.12    \\
   5135.1400 & 59.01 & 45.7&   0.008&   0.950 &     1.12    \\
   5219.0450 & 59.01 & 58.9& --0.052&   0.795 &     1.18    \\
   5220.1079 & 59.01 & 92.5&   0.298&   0.796 &     1.18    \\
   5322.7710 & 59.01 & 83.6& --0.318&   0.483 &     1.38    \\
   6017.7990 & 59.01 & 31.0& --0.257&   1.112 &     1.27    \\
   6025.7170 & 59.01 & 35.2&   0.067&   1.440 &     1.34    \\
   4917.3800 & 60.01 & 60.9& --1.440&   0.559 &     2.43    \\
   4943.8990 & 60.01 & 76.2& --1.640&   0.205 &     2.44    \\
   4947.0200 & 60.01 & 50.8& --1.130&   0.559 &     2.00    \\
   4949.0110 & 60.01 & 33.6& --1.450&   0.630 &     2.15    \\
   4998.5410 & 60.01 & 83.4& --1.100&   0.471 &     2.23    \\
   5063.7200 & 60.01 & 65.1& --0.620&   0.976 &     2.06    \\
   5089.8320 & 60.01 & 78.8& --1.160&   0.205 &     1.97    \\
   5092.7900 & 60.01 &146.6& --0.610&   0.380 &     2.36    \\
   5181.1690 & 60.01 & 74.8& --0.599&   0.859 &     2.02    \\
   5182.5900 & 60.01 & 78.7& --0.910&   0.745 &     2.25    \\
   5212.3600 & 60.01 &132.6& --0.960&   0.205 &     2.33    \\
   5255.5056 & 60.01 &161.6& --0.670&   0.205 &     2.40    \\
   5272.0460 & 60.01 & 42.0& --1.070&   0.986 &     2.23    \\
   5276.8690 & 60.01 & 82.8& --0.440&   0.859 &     1.93    \\
   5306.4600 & 60.01 & 42.6& --0.970&   0.859 &     2.01    \\
   5310.0400 & 60.01 & 23.4& --0.980&   1.137 &     1.97    \\
   5311.4500 & 60.01 &115.6& --0.420&   0.986 &     2.39    \\
   5356.9700 & 60.01 & 95.6& --0.280&   1.264 &     2.31    \\
   5361.4670 & 60.01 &155.0& --0.370&   0.680 &     2.49    \\
   5385.8880 & 60.01 & 77.5& --0.819&   0.742 &     2.13    \\
   5385.8880 & 60.01 & 87.7& --0.820&   0.742 &     2.24    \\
   5416.3740 & 60.01 & 62.8& --0.930&   0.859 &     2.20    \\
   5421.5510 & 60.01 & 76.7& --1.330&   0.380 &     2.26    \\
   5431.5200 & 60.01 & 86.5& --0.469&   1.121 &     2.25    \\
   5442.2640 & 60.01 & 69.1& --0.910&   0.680 &     2.07    \\
   5447.5550 & 60.01 & 57.1& --1.050&   1.044 &     2.44    \\
   5449.2200 & 60.01 & 50.4& --0.690&   1.264 &     2.22    \\
   5485.7000 & 60.01 &104.3& --0.120&   1.264 &     2.23    \\
   5508.3980 & 60.01 & 60.8& --1.230&   0.859 &     2.47    \\
   5533.8200 & 60.01 & 70.2& --1.230&   0.559 &     2.27    \\
   5548.4500 & 60.01 & 68.4& --1.270&   0.550 &     2.28    \\
   5550.0800 & 60.01 & 46.3& --0.980&   1.228 &     2.42    \\
   5595.8020 & 60.01 & 31.5& --1.530&   0.859 &     2.38    \\
   5603.6500 & 60.01 & 33.5& --1.690&   0.380 &     2.09    \\
   5620.5943 & 60.01 & 71.5& --0.310&   1.545 &     2.35    \\
   5625.7300 & 60.01 & 36.2& --1.120&   0.933 &     2.12    \\
   5659.7750 & 60.01 & 27.1& --0.520&   1.600 &     2.03    \\
   5702.2400 & 60.01 & 90.1& --0.880&   0.745 &     2.30    \\
   5726.8300 & 60.01 & 78.9& --0.810&   1.044 &     2.42    \\
   5740.8600 & 60.01 & 72.6& --0.530&   1.160 &     2.19    \\
   5742.0900 & 60.01 & 42.6& --0.830&   1.091 &     2.07    \\
   5743.1925 & 60.01 & 36.5& --0.730&   1.282 &     2.07    \\
   5744.7770 & 60.01 & 43.9& --1.030&   0.986 &     2.18    \\
   5865.0270 & 60.01 & 29.1& --0.830&   1.410 &     2.17    \\
   6031.2700 & 60.01 & 37.0& --0.740&   1.282 &     2.07    \\
   6183.9000 & 60.01 & 45.4& --0.920&   1.160 &     2.24    \\
   6330.1510 & 60.01 & 27.5& --0.720&   1.773 &     2.37    \\
   6382.0600 & 60.01 & 43.2& --0.750&   1.436 &     2.31    \\
   6385.1870 & 60.01 & 72.0& --0.360&   1.600 &     2.41    \\
   6637.9600 & 60.01 & 45.0& --0.320&   1.773 &     2.22    \\
   6650.5200 & 60.01 & 40.4& --0.110&   1.953 &     2.13    \\
   6737.7600 & 60.01 & 39.5& --0.670&   1.600 &     2.32    \\
   5294.1133 & 60.02 & 78.8& --0.690&   0.000 &     2.12    \\
   4815.8000 & 62.01 &127.1& --0.820&   0.185 &     1.94    \\
   4844.2088 & 62.01 & 94.0& --0.890&   0.278 &     1.73    \\
   4948.6300 & 62.01 & 39.3& --0.950&   0.544 &     1.42    \\
   4961.9400 & 62.01 & 74.6& --1.090&   0.434 &     1.87    \\
   5052.7500 & 62.01 & 55.0& --0.160&   1.375 &     1.66    \\
   6731.8100 & 62.01 & 30.3& --0.739&   1.166 &     1.58    \\
   6645.0940 & 63.01 & 60.9&   0.120&   1.380 &     0.77    \\
   6221.8900 & 71.01 & 55.4& --0.760&   1.542 &     0.83    \\
   5989.0450 & 90.01 & 50.2& --1.414&   0.189 &     1.12    \\
\hline  
\label{EWs2}
\end{longtable}         
  
\end{appendix}          

\end{document}